\documentclass[floatfix,plainnat,tightenlines,prd,nofootinbib,notitlepage,superscriptaddress,10pt,twocolumn]{revtex4-1}

\pdfoutput=1

\usepackage{float}
\restylefloat{table}

\usepackage{mathrsfs}
\usepackage{amsmath}
\usepackage{amssymb}
\usepackage{color}
\usepackage{graphicx}
 \usepackage{notes2bib}
 \usepackage{natbib}
\usepackage{bibentry}

\begin{document}
\title{Negative votes to depolarize politics}
\author{Karthik H. Shankar}
\email{Email: kshankar79@gmail.com}
\affiliation{Center for Memory and Brain, Boston University}

\begin{abstract}

The controversies around the 2020 US presidential elections certainly casts serious concerns on the efficiency of the current voting system in representing the people's will. Is the naive Plurality voting  suitable in an extremely polarized political environment?  Alternate voting schemes are gradually gaining public support, wherein the voters rank their choices instead of just voting for their first preference. However they do not capture certain crucial aspects of voter preferences like disapprovals and negativities against candidates. I argue that these unexpressed negativities are the predominant source of polarization in politics. I propose a voting scheme with an explicit expression of these negative preferences, so that we can simultaneously decipher the popularity as well as the polarity of each candidate. The winner is picked by an optimal tradeoff  between the most popular and the least polarizing candidate. By penalizing the candidates for their polarization, we can discourage the divisive campaign rhetorics and pave way for potential third party candidates. 
 
\end{abstract}

\maketitle{}

Social choice theorists have pondered over alternate voting systems for more than two centuries since Nicolas de Condorcet \cite{condorcet1785essay}. The ranked voting systems, wherein the voters rank their choices instead of just voting for their first preference, have been thoroughly explored and their deficiencies have been mathematically nailed down. 
Arrow's Impossibility theorem \cite{arrow1950difficulty} proves that it is impossible to always pick a winner in a ranked voting system that satisfies certain basic intuitive criteria. Further, the Gibbard-Satterthwaite (GS) theorem \cite{gibbard1973manipulation,satterthwaite1975strategy} proves that in any ranked voting system, some voters can always strategically misrepresent their votes to alter the results, which implies  an impossibility of a \emph{strategy-proof} voting system. So, it appears that there cannot exist a voting system that fairly represents the voter preferences, prompting the notion that a perfect democracy is mathematically impossible.\footnote{The Stanford Encyclopedia of Philosophy provides a good nontechnical introduction to various voting systems. 
https://plato.stanford.edu/entries/voting-methods/ }  

It may very well be true that we cannot design a perfect voting system, nevertheless we can certainly do much better than the naive \emph{plurality voting} system wherein the voters only vote for their first preference. Any alternate voting method is of course more complex, but the advent of electronic voting vastly simplifies its implementation. So there is no valid excuse to not facilitate smarter and fairer elections. 

The major problem with the plurality voting system is that when two strong candidates emerge with certain critical mass of support prior to the election, then rest of the support will automatically coagulate around them, simply because many voters don't want to waste their votes on a third candidate. This is very deceptive because many of those votes are not direct support for the respective candidates, rather they are anti-votes against the opposing candidate. To make matters worse, the candidates understand this phenomenon and indulge in some divisive campaign rhetorics to collect more of their opponent's anti-votes. Many voters would thus strategically refrain from voting for their real preference (the third party candidate) because their priority is to defeat one of the top two contenders by voting for the other. 
Consequently, the third party candidates are stifled out of competition. 

Ranked voting systems alleviate this issue significantly because the voters have the opportunity to elaborate their preferences in a more detailed fashion. The voters can express their dislike for a candidate by ranking them the last, however this is not the same as explicitly casting a negative vote. The only way to prevent the anti-votes for a candidate from being masked into votes for another candidate is to explicitly express them on the ballot as negative votes. 

Moreover, it is well established that people's choice very much depends on how the decision problem in front of them is framed \cite{tversky1981framing}\footnote{For example, people make completely different choices when a problem is framed as an increase in monetary gain as opposed to avoiding monetary loss.}. By not allowing the voters to freely express their negative preferences, we are essentially ill-framing the decision problem and needlessly tampering with voter's psychology.  
Here I shall propose a voting system which explicitly incorporates negative votes. 

\textbf{Normed Negative Voting : }
Consider a voting system where each voter assigns a positive or negative number to each candidate such that the magnitudes of all the numbers sum up to 10. 

Example : In a three candidate election amongst $ \{ A, B, C \} $, voter-1 could vote as $ \{+7, -1, -2 \}$ while voter-2 could vote as $\{ +3, +2, -5 \} $. Both the voters express the same rank ordering  of preferences, namely  $A > B > C$, however their ballots clearly contain far richer information than just the relative ranks. Requiring the sum of magnitudes of the vote to be a constant (10) is mathematically termed as \emph{Normed}; it serves to ensure the rule of equality \emph{one-person-one-vote}.    
 
\textbf{Vote aggregation : } 
For each candidate, aggregate the positive votes from all voters as P, and aggregate the negative votes from all voters as N.  Then define for each candidate \emph{Popularity} $\equiv$ P-N, and \emph{Polarity} $\equiv$ N/P.  An absolutely noncontroversial candidate who does not acquire any negative votes will have the lowest polarity of zero. On the other hand, a candidate with almost equally large positive and negative votes is by definition extremely polarizing with a polarity of 1. For obvious reasons, candidates with polarity larger than 1 should be disqualified.

\textbf{Winning metric : } 
To determine the winner, we shall construct a metric W that rewards popularity and penalizes polarity. It has to be a monotonically increasing function of popularity and a monotonically decreasing function of polarity.  The candidate with the highest value of W is the winner.  If two candidates have equal popularity, the one with lower polarity will have to win. Consider the following metric parametrized by two positive constants $(c,b)$.
  \begin{equation}
  W_{b}^{c} (P,N) \equiv \frac{P- c N}{1+ b \,N/P}
  \label{win}
  \end{equation}

Let us first examine this function with $c=b=1$.  Consider an election with three candidates and two voters as shown below.  
\begin{table}[H]
\begin{center}
\begin{tabular}{|c||c|c||c|c|c|c|c|}
\hline
Candidate & voter-1 & voter-2 &P & N & P-N & N/P & $W_{1}^{1}$  \\
 \hline 
\hline
A  & 10 & -5  &  10 & 5 &  5 & 0.5  & 3.33 \\
\hline
B & 0 & 4  & 4 & 0 & 4  & 0   & 4 \\
\hline
C & 0 & 1  & 1 & 0 & 1  & 0   & 1 \\
\hline
\end{tabular}
\end{center}
\end{table}

Although A has the highest popularity, B still wins over A because of lower polarity. This seems to be a fair result for this election. However, we should note that this voting scheme is severely prone to strategic-voting. Voter-1 has naively expressed a clear preference for A and no dislike for B. On the other hand, voter-2 might  simply prefer B over A  (and no real dislike for A), however by misrepresenting the preference and casting a negative vote to increase the polarity of A, voter-2 can enhance the chances of B's victory. Game-theoretically speaking,  voter-1 would foresee this and in turn misrepresent his preference to include a negative vote for B.
\footnote{When there are  more than two candidates it is difficult to strategically cast negative votes without hampering the positive votes for the preferred candidate, which could inhibit the winning chances of the preferred candidate.}

\textbf{Avoid over-penalization of Negative Votes : }
Strategically misrepresenting the negative votes as described above should be discouraged. It cannot be completely eliminated, but it can be suppressed by appropriately constraining the winning metric.  We should ensure that the winning metric does not over-penalize the negative votes because that would give the voters an opportunity to exploit it. In order to quantify what exactly we mean by ``over-penalize", let us first note that plurality voting method is indeed the optimal voting scheme in a two-candidate election. So we shall demand that a perfect preference for a candidate by one voter cannot be overridden by another voter's negative vote. To be more specific, 
\begin{itemize}
\item{\emph{In a 2-candidate election  with just two voters, if voter-1 gives a perfect preference for one candidate, then that candidate cannot loose the election regardless of how voter-2 votes. }}
\end{itemize}
To work out this constraint, let voter-1 cast  +10 votes for A and voter-2 cast a  negative vote -X for A as shown below.    
 \begin{table}[H]
 \begin{center}
\begin{tabular}{|c||c|c||c|c|c|}
\hline
Candidate &   voter-1 & voter-2 & P-cN & N/P & $W_{b}^{c}$  \\
 \hline 
\hline
A  & 10 & -X  & 10-cX & $X/10$  & $ 10 \left( \frac{10-cX}{10+bX} \right) $ \\
\hline
B & 0 & 10-X  & 10-X  & 0   & 10-X \\ 
\hline
\end{tabular}
\end{center}
\end{table}
For any value of $X$ between 0 and 10, we demand that candidate B should not be able to win. 
\begin{eqnarray}
&&  \left( \frac{10-cX}{1+b\,X/10} \right) \ge (10-X)   \\
& & \Rightarrow   \frac{c+b-1}{b}  \le X/10   \label{V2} \\
& &\Rightarrow c+b \le1
\end{eqnarray}
Since B cannot win anyway, the best option for voter-2 is to choose $X=0$, so that A and B are tied up. 

We can now generalize the constraint for $m$ candidates and a large number of voters. Again, let voter-1 assign +10 votes to A, and voter-2 assign -X votes to A. The reminder of 10-X votes of voter-2 can be assigned as positive votes for any of the other $(m-1)$ candidates. Here voter-2 should be viewed as a statistical representative of all voters who voted -X for A; and all these voters are assumed to be independent minds uninfluenced by each other. On average, each of the other candidates would have received $(10-X)/(m-1)$ votes from voter-2. So, the above constraint can be reframed at a statistical level as  
\begin{eqnarray}
&& \left( \frac{10-cX}{1+b\,X/10} \right) \ge \frac{(10-X)}{m-1} \,\, \Rightarrow  \nonumber \\
&&  m-2 \ge [ (m-1)c +b -1] X/10 - b (X/10)^2   \label{Vm} 
\end{eqnarray}

\begin{figure} 
\includegraphics[width=0.46 \textwidth]{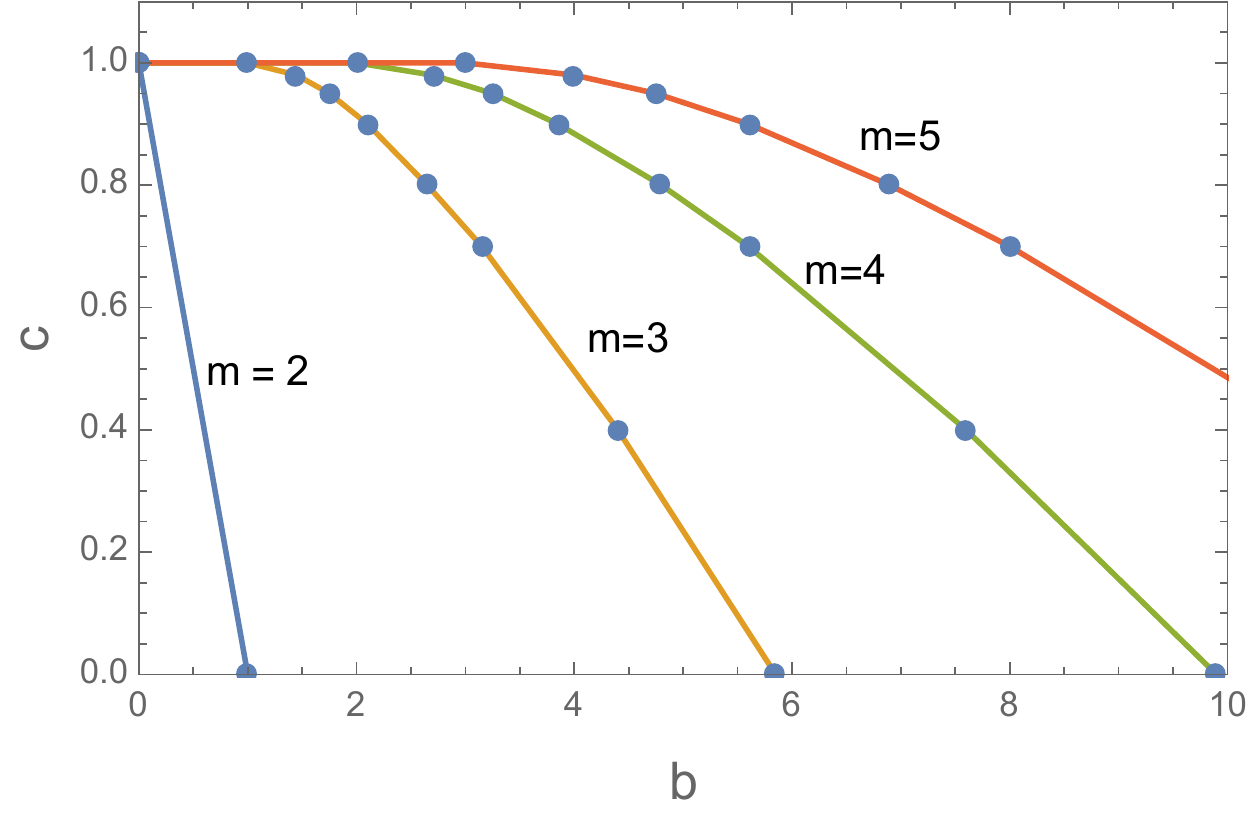}
\caption{Maximal-Penalty metrics}
\label{cbplot}
\end{figure}

In eq.\ref{V2} the r.h.s attains a minimum at X=0, which is where the equality should be implemented. But that is not true in eq.\ref{Vm} when $m>2$. The acceptable range of parameters (c,b) for which eq.\ref{Vm} holds for all values of X is plotted in fig.\ref{cbplot} for $m=2,3,4,5$; the region under the curves corresponding to each $m$ contain the admissible values for the parameters. First we note the obvious, that $c$ cannot be greater than 1 because negative votes shouldn't weigh more than positive votes. For $m=2$, note that $c=1$ is admissible only with $b=0$. But for $m>2$,  $c=1$ is admissible for any $b \le m-2$. In particular, note that $(c=1,b=1)$ is acceptable for all $m>2$.  By choosing the parameters on the curves of fig.\ref{cbplot}, we are hitting the limit of over-penalization of negative votes; so these curves are essentially the \emph{maximal-penalty} metrics.

Here we have only analyzed the winning metric function of the form  eq.~\ref{win}. But it is fairly straightforward to imagine other functional forms that introduce nonlinearities to penalize the negative votes more adversely, like
\begin{equation}
 \frac{(P-N)}{e^{N/P}},  \,\,\,  \frac{(P-N)^2}{(P+N)}, \, \,\,   (P-N)^{1-N/P}.
 \label{altmetrics}
\end{equation}
The first two are acceptable, but the third is not acceptable because the metric function must overall be linear in popularity so that its properties do not depend on the size of the electorate. 
These functions can be analyzed in a procedure similar to that discussed above with introduction of some free parameters (analogous to $c,b$)  to ensure that negative votes are not over-penalized. The maximal-penalty metrics thus obtained show similar qualitative behavior as shown in fig.\ref{cbplot}, so there is not much utility to further discuss these alternate functional forms of metric.

\section*{Maximizing Voter satisfaction}

  Let's denote the voters by latin indices $i,j,k...$ and the candidates by greek indices $\mu, \nu, \alpha...$. Let's denote the positive votes from  voter-$i$ to candidate-$\mu$ as $p^{i}_{\mu}$, and the negative votes as $n^{i}_{\mu}$.  Only one of the two, $p^{i}_{\mu}$ or $n^{i}_{\mu}$, will be nonzero for a specific $i$ and $\mu$.
Summing over all the voters will give the net positive votes $P_{\mu}= \sum_{i} p^{i}_{\mu}$ and  net negative votes $N_{\mu}= \sum_{i} n^{i}_{\mu}$ gathered by each candidate-$\mu$. 

If candidate-$\alpha$ is declared the winner of the election, we define can define the voter-satisfaction  to be 
 \begin{eqnarray}
&& s^{i}_{\alpha} = p^{i}_{\alpha}-n^{i}_{\alpha}+ \sum_{\mu}^{ \mu  \neq \alpha} n^{i}_{\mu}  
  \nonumber \\
&&S_{\alpha} = \sum_{i} s^{i}_{\alpha} = P_{\alpha} - N_{\alpha}  + \sum_{\mu}^{ \mu  \neq \alpha}  N_{\mu}  
 \label{satisfaction} 
 \end{eqnarray} 
 \begin{equation}
\bar{S}_{\alpha} =S_{\alpha}  - \left[ \sum_{\mu}^{ \mu  \neq \alpha}  P_{\mu} \right] 
\label{Sbar}
\end{equation}

 The first two terms in the r.h.s of eq.~\ref{satisfaction} correspond to the satisfaction/dissatisfaction explicitly triggered by the winning candidate, while the third term represents the satisfaction triggered by all the loosing candidates. Negative votes are directly responsible for a candidate to loose the election, and so those negative votes assigned to loosing candidates can be deemed to have satisfactorily performed their job, and hence contributes to the voter satisfaction as the third term in the r.h.s of eq.~\ref{satisfaction}.
 
 One might be tempted to include the dissatisfaction due to the positive votes accrued by the loosing candidates as shown in eq.~\ref{Sbar}. But notice that it is a measure of \emph{inaction} of certain positive votes that failed to deliver victory. It should be treated simply as a lack of satisfaction that was potentially attainable, rather than a negative quantity to be subtracted from voter satisfaction. While defining the voter satisfaction, it is critical to acknowledge the intrinsic asymmetry between the positive and negative votes in the voter's psychology-- \emph{positive votes succeed only when they are cast to the winner, but negative votes succeed when they are cast to any loser.}  
 
Ideally we would like the election outcome to maximize the satisfaction of all voters, \emph{i.e.} $ S_{\alpha}$ should be maximum for the winning candidate-$\alpha$. Notice that if we only considered the first two terms in the r.h.s of eq.\ref{satisfaction}, then we simply have to maximize the popularity given by $(P_{\alpha} -N_{\alpha})$, which is exactly what the winning metric $W_{0}^{1}$ would implement. It is however not obvious whether some other winning metric will always yield the winner who maximizes voter satisfaction. Consider an example of a 4-candidate election with aggregated positive and negative votes  as shown below.    

\begin{table}[H]
\begin{center}
\emph{Election-0}

\begin{tabular}{|c||c|c||c||c|c|c|c|}
\hline
Candidate &   P& N& $S$  & $W_{0}^{1}$ &  $W_{1}^{1}$  & $W_{2}^{1}$  & $W_{0.5}^{0.5}$\\
 \hline 
\hline
A  & 11 & 5  &  10  & 6.0 & 4.12 & 3.14  &6.92 \\
\hline
B & 7 &  3   & 10  & 4.0 & 2.8 & 2.15 & 4.53 \\
\hline
C & 6 & 1  & 13  & 5.0 & 4.28 & 3.75 & 5.08\\
\hline
D & 3 & 0  &  12  & 3.0 & 3.0 & 3.0  &3.0\\
\hline
\end{tabular}
\end{center}
\end{table}
Here, candidate-C has the maximum voter satisfaction, and also wins under the metrics $W_{1}^{1}$ and  $W_{2}^{1}$. But this may not always happen; in some elections the winning metrics may not yield the candidate who maximizes voter satisfaction. To understand how reliably these metrics correlate to maximal voter satisfaction, we shall simulate a large number of $m$-candidate elections with randomly distributed $(P_{\mu},N_{\mu})$ with at least one  candidate qualified to win. Then we calculate the probability that a  winning metric yields a winner who maximizes the voter satisfaction. Table \ref{sims} shows the results for four different metrics  and various values of $m$.

 \begin{table}[H]
\begin{center}
\begin{tabular}{|c|c|c|c|c|}
\hline
  & $W_{0}^{1}$ &  $W_{1}^{1}$  & $W_{2}^{1}$ & $W_{0.5}^{0.5}$ \\
 \hline 
\hline
$m=3$  &87\% &91\% & 86\%  & 80\% \\
\hline
$m=4$   &84\% &89\% &87\%& 76\% \\
\hline
$m=5$  &82\% &88\% &88\% & 73\%\\
\hline
$m=8$   &79\% &88\% &91\% & 70\% \\
\hline
$m=20$   &79\% &91\% &93\% & 70\% \\
\hline
\end{tabular}
\end{center}
\caption{Winning  Metrics Correlation to Maximal Voter Satisfaction}
\label{sims}
\end{table}

Interestingly, the metrics $W_{1}^{1}$  and $W_{2}^{1}$ perform better in aligning with  voter satisfaction for larger values of $m$, while this effect is not seen in  $W_{0}^{1}$. This suggests that the effect of polarity in the functional form of the metric (denominator of eq.~\ref{win}) truly captures the voter satisfaction in a very non-obvious way.  This is indeed nontrivial because the metric is a function of only the votes obtained by any particular candidate, but voter satisfaction is a function of votes obtained by all candidates. It is also very clear from the performance of $W_{0.5}^{0.5}$ that a metric with $c<1$ is suboptimal. The most suitable metric that remains within the maximal penalty limit and  works for all $m \ge 3$ is the metric $W_{1}^{1}$. 

If we altered our definition of voter satisfaction to be $\bar{S}_{\alpha}$ in eq.~\ref{Sbar}, we find that the winner picked by $W_{0}^{1}$ almost always maximizes voter satisfaction. So, whether or not the metric function should penalize polarity (eq.\ref{win}) very much depends on our definition of voter satisfaction.  

More generally, we could ask the following question-- if the aim is to pick the candidate who would maximize the voter satisfaction, then why do we need to pick the winner in a round-about manner using a winning metric, rather than  directly picking the candidate maximizing the voter satisfaction?  This is because the prescription of a winning metric a priori informs the voters how much their negative votes weigh against their positive votes. Without a winning metric we wouldn't have a tool to discourage the voters from strategic voting.

 \section*{Comparison with Ranked Voting}

It is straightforward to convert the normed negative votes into ranked votes, as long as voters cast distinct votes to various candidates. For example, consider the 4-candidate election with three voters as shown below. Candidate-B is the winner according to the metrics $W_{0}^{1}$, $W_{1}^{1}$ and the voter satisfaction. 

\begin{table}[H]
\begin{center}
\emph{Election-1}
\begin{tabular}{|c||c|c|c||c||c|c|c|}
\hline
Candidate &   voter-1 & voter-2 &  voter-3 & Borda  & $W_{0}^{1}$  & $W_{1}^{1}$ &  S \\
 \hline 
\hline
A  & 5 [1] & -5 [4] & 3 [2]  & 3+0+2 & 3.0 &1.84 & 6\\
\hline
B & 2 [2] & 0 [3] & 4 [1]   & 2+1+3  & 6.0 & 6.0 & 14\\
\hline
C & 1 [3] & 1 [2] & -2 [4]  & 1+2+0  & 0.0 & 0.0 & 6 \\
\hline
D & -1 [4] & 4 [1] & 1  [3] & 0+3+1 & 4.0 & 3.33 & 11\\
\hline
\end{tabular}
\end{center}
\end{table}
 The rank corresponding to each vote is expressed in square brackets next to the vote. Let us now bypass the vote itself and only consider the ranks. There are different ways to pick a winner in a ranked voting system. In \emph{Condorcet} method \cite{fishburn1977condorcet}, we split the $m$-candidate race into a bunch of 2-candidate head-to-head competitions. The winner is the candidate who wins every head-to-head competition.  In the above example, voter-1 ranks A higher than B while voter-2 and voter-3 rank B higher than A; hence B defeats A. Similarly B defeats C and D in head-to-head competitions.Thus candidate-B is the Condorcet winner in this example. However, there are many situations where this procedure does not yield a clear winner, and we end up in a rock-paper-scissor configuration of cyclic loop of winners.  

In \emph{Instant Runoff}, votes are counted in multiple stages. In each stage, the candidates with the least number of rank-1 votes get eliminated, and subsequently the ranks of all other candidates are boosted up by one in those ballots that had the eliminated candidate in the rank-1 position. So, the voter's ballot is not eliminated after the first preference is eliminated, instead the subsequent  preferences are considered in order. In the above example, candidate-C gets eliminated in the first round with zero rank-1 votes, and  A, B \& D end up in a tie. Such a tie situation is unlikely when there are many voters, and the counting will proceed to next stage of elimination. For instance if there were two voters who vote like voter-3, then A and D get eliminated in the second stage with just one rank-1 vote each, leaving candidate-B the winner.  

A major issue with Instant Runoff is that it does not satisfy the basic criterion of \emph{monotonicity}. That is, a winning candidate can become a looser by getting ranked higher by a voter, which is intuitively strange and unacceptable. It happens because a stronger candidate survives an earlier stage of elimination due to the rank modifications. This gives a lot of leeway for the voters to strategically attempt to eliminate strong opposition candidates at earlier stages of elimination, rather than vote according to their innate preferences. It is of course very difficult to strategize for multistage elimination with a large number of voters, but there is nothing stopping the voters from attempting to strategize. Any procedure which calls for multistage elimination of candidates is prone to this issue.         
The GS theorem \cite{gibbard1973manipulation} shows that it is not possible to totally prevent strategic voting, but we can curtail it significantly by satisfying the monotonicity criterion.

 \emph{Borda} proposed a metric that aggregates a weighted sum of the ranks accrued by each candidate \cite{borda1784memoire}. In an $m$-candidate race, rank-1 gets a weight $(m-1)$, rank-2 gets a weight $(m-2)$ and so on, with rank-m ending at weight zero. In the above example, candidate A has accrued one rank-1, one rank-4 and one rank-2, yielding a net Borda count of 3+0+2 =5; while B attains a Borda count of 6 and wins the election.   Borda metric satisfies the monotonicity criterion, but it violates another intuitive criterion, \emph{Independence of Irrelevant Alternatives} (IIA).
  
The IIA criterion requires that  if the voters were allowed to modify their votes without changing the relative preference between a winning candidate and a specific looser, then that looser shouldn't be able to win due to the vote modifications. 
Suppose candidate-$\alpha$ is the winner and is ranked higher than a loosing candidate-$\mu$ by some voter, then  no matter how the voter changes the ranks while holding $\alpha$ higher than $\mu$, there should be no chance for candidate-$\mu$ to win. 
The Condorcet method of counting the head-to-head competitions only cares about the relative rankings, hence it will satisfy IIA criterion. Any other method that weighs the absolute rank positions, like the Borda metric, will violate this criterion. The IIA criterion seems overrated for its intuitiveness, primarily because it is a necessary condition for Arrow's impossibility theorem  which proves that it is not always possible to find a winner in a ranked voting election when certain intuitive conditions are met \cite{arrow1950difficulty,fishburn1970arrow, reny2001arrow}.   

The normed negative voting (NNV) method in this article would not satisfy the IIA criterion, because these votes are cardinal in nature, which represents much richer information than relative ranking. Hence it is not subject to implications of Arrow's theorem \cite{arrow1950difficulty}. Furthermore, the NNV also violates the monotonicity criterion.  
If a voter increases the positive votes for the winner, then obviously the winning metric value as calculated by eq.~\ref{win} would further increase. But any increase in positive votes for the winning candidate must be compensated by a decrease in votes to other candidates, as per the normed voting rules. Let's suppose that the negative votes to another candidate is reduced. It is now possible for this candidate to emerge as the new winner, if $c+b>1$. This violates the monotonicity criterion. However, notice that this is precisely what we prevented at a statistical level, by restricting the metric to stay under maximal penalty (eq.~\ref{Vm}). Although the monotonicity criterion does not hold at individual instances, it holds on average with large number of voters. That is why it is not possible to strategically exploit the negative votes without large scale coordinated effort among voters. 

In \emph{Approval voting} method, the voters can approve of any number of candidates and disapprove of the others, and every approved candidate gets +1 vote. In a way, this violates the basic principle of \emph{one-person-one-vote}. But we can rectify the method by assigning  -1 vote to every disapproved candidate, thereby revising it as normed-Approval voting.  We can view this as a special case of the NNV procedure with a uniform magnitude of positive or negative vote for every candidate, which is clearly an unnecessary restriction on voter expression.  

The NNV method gathers the voter preferences in a very rich format. To emphasize this, let's modify the votes of voter-2 in the previous example of \emph{Election-1} without affecting the relative ranks, as shown below.

\begin{table}[H]
\begin{center}
\emph{Election-2}
\begin{tabular}{|c||c|c|c||c||c|c|c|}
\hline
Candidate &   voter-1 & voter-2 &  voter-3 & Borda  &  $W_{0}^{1}$ & $W_{1}^{1}$ &  S \\
 \hline 
\hline
A  & 5 [1] & -1 [4] & 3 [2]  & 3+0+2 & 7.0 & 6.2 & 10\\
\hline
B & 2 [2] & 0 [3] & 4 [1]   & 2+1+3  & 6.0 & 6.0 & 10\\
\hline
C & 1 [3] & 1 [2] & -2 [4]  & 1+2+0 & 0.0 & 0.0  & 2 \\
\hline
D & -1 [4] & 8 [1] & 1  [3] & 0+3+1 & 8.0 &7.2 & 11\\
\hline
\end{tabular}
\end{center}
\end{table}
 Candidate-B is still the winner under all the ranked voting methods discussed above. But candidate-B looses to both A and C  under NNV metrics $W_{0}^{1}$, $W_{1}^{1}$,  and the voter satisfaction.  \emph{Election-1} is identical to \emph{Election-2} under ranked voting methods, but very different under NNV. 
This clearly illustrates the effect of superficially considering the ranks while ignoring the deeper negative preferences of voters. It is also not difficult to imagine the frustration that would develop among the voters if their votes aligned with \emph{Election-2}, but ranked voting method results in B's victory, which is aligned with \emph{Election-1}.

\section*{Conclusion}

I have discussed the importance of  negative votes in  capturing the voter preferences in a richer  format.  To prevent voters from exploiting the negative votes for strategic voting, we constrained the winning metric  to avoid over-penalization of the negative votes. 
Since the negative votes for the disliked candidates come at the expense of positive votes for their preferred candidates, the voters are incentivized to vote according to their true preference, suppressing the intent for strategic voting. 

This would also have a serious impact on how  election campaigns are conducted. For the fear of accumulating negative votes, the candidates would refrain from divisive rhetorics and stay focused on constructive issues. The major parties cannot afford to nominate a polarizing candidate because they understand that it would directly pave way for a third party victory. The whole political arena can thus be depolarized.

The candidates will  stay well-behaved by conducting decent campaigns and the voters will stay well-behaved by voting their true preferences.
We can then hope for a future where no candidate receives any negative vote. --\emph{Dream of an Ideal society}.


%

\end{document}